\documentclass{kapproc} 

\usepackage{procps} 

\usepackage[dvips]{graphicx}
\usepackage{epsfig}

\upperandlowercase

\normallatexbib 

\begin{document}


\articletitle{Sum rules \MakeLowercase{and} energy scales 
\MakeLowercase{in} B\MakeLowercase{i}S\MakeLowercase{r}C\MakeLowercase{a}C\MakeLowercase{u}O}


\author{A.F. Santander-Syro, R.P.S.M. Lobo, N. Bontemps}
\affil{Laboratoire de Physique du Solide (UPR5 CNRS) ESPCI\\
10 rue Vauquelin, 75231 Paris cedex 05, France}
\email{nicole.bontemps@espci.fr}


\begin{abstract}
From very high accuracy reflectivity spectra, we have derived the optical 
conductivity and estimated the spectral weight up to various cut-off frequencies 
in underdoped Bi$_2$Sr$_2$CaCu$_2$O$_{8+\delta}$ (Bi-2212). 
We show 
that, when evaluating the 
optical spectral weight over the full conduction band (1 eV), 
the kinetic energy decreases in the superconducting state, unlike in 
conventional BCS superconductors. As a consequence, the Ferrell-Glover-Tinkham sum rule
 is not satisfied up to this energy scale.
This stands as a very unconventional behavior, contrasted with 
the overdoped Bi-2212 sample.
\end{abstract}


\section*{Introduction}
A long lasting debate about the cuprate superconductors, stems from the differences 
and similarities with BCS behavior. The actual pairing mechanism, 
which eventually results in lowering the 
free energy of the system, is not yet known. In BCS superconductors, 
the free energy gain results from a competition of electronic kinetic energy  increase  
and an eventually larger potential energy decrease.
 If the free energy in cuprates can be
(somewhat arbitrarily) separated between potential and kinetic energy \cite{Chakra, Chubu}, 
then the latter is  measured by optics \cite{BasovVDM,NormanPepin}. 
The kinetic energy can be inferred from 
the partial sum rule  or spectral weight $W$, defined as:

\begin{equation}
  W(\Omega) = \int_0^\Omega \sigma_1(\omega,T) d\omega
  \label{eq:DefW}
\end{equation}
where $\sigma_1(\omega,T)$ is the frequency ($\omega$) and
temperature ($T$) dependent conductivity, and $\Omega$ is a
cut-off frequency. Setting $\Omega = \Omega_B$, where $\Omega_B$
is the conduction band widht, one can get the kinetic energy 
$E_k$ per copper site \cite{NormanPepin}, through:

\begin{equation}
  W(\Omega_B) = {\pi \over {2}}{e^2 \over {\hbar ^2} }{a^2 \over {V_u} } {[-E_k]}
  \label{eq:DefEk}
\end{equation}
where $a$ is the lattice parameter, and $V_u$ the volume per Cu site.

In the superconducting state, the integral in eq.1 or 2 includes the contribution of 
 the superfluid, i.e. the weight of the $\delta(\omega)$ function centered  at 
 zero frequency.  The Ferrell-Glover-Tinkham (FGT) sum rule
\cite{FGT-1,FGT-2} requires that the spectral weight 
lost at finite frequency in
the superconducting state must be retrieved in the spectral
weight $W_s$ of the $\delta$ function. In conventional superconductors, it was 
found to be fulfilled if integrating up to $\hbar \Omega_0 \sim 4 \Delta$ ($\Delta$
is the superconducting gap). $\hbar \Omega_0$ is a characteristic energy of the boson spectrum
responsible for the pairing mechanism. The FGT sum rule would then be exhausted for cuprates, 
if conventional, for  $\hbar \Omega_0 \sim  0.1$~eV (assuming $\Delta \sim$ 25~meV) 
\cite{GapHTc-1,GapHTc-2}. 

The studies of the FGT sum rule, 
first performed from c-axis (interlayer) optical conductivity data, showed
a strong violation, interpreted
as  a change of interlayer kinetic energy \cite{BasovVDM,Basov}. However, 
the amount of kinetic energy saving which was found is too small to account for the
condensation energy. Although
 {\it in-plane} data appeared firstly to yield a conventional behavior \cite{Basov}, 
 these early results were subsequently contradicted by
ellipsometric and infrared data \cite{vdmScience}. Our own infrared-visible 
reflectivity experiments, in Bi-2212, showed that in-plane spectral weight lost from  
the visible range is transferred into the $\delta$ function \cite{AFSS-EPL}.
These two independent sets of data yielded a {\it decrease} of kinetic energy below $T_c$ 
of the order of 1~meV.

 The present paper implements our previous report \cite{AFSS-EPL} by using the 
 {\it partial sum rule} in Eq.2.
 Although this is in principle equivalent to the  FGT sum rule, we found 
 this method to be more robust, because we can trace the entire temperature evolution of 
 the spectral weight.  We focus here on  the underdoped thin film from the Bi-2212 family.
We pin down the raw reflectivity data which allows to establish small changes at high energy (up to 
10000~cm$^{-1}$) in the conductivity, hence in 
the spectral weight, thus illustrating  why our unprecedented resolution is a necessary condition 
to trace this phenomenon. 
Our present, more elaborate analysis confirms that within  error bars,
the in-plane kinetic energy, calculated from Eq.2, decreases in the superconducting state. Using 
the partial sum rule in Eq.2 (and not only the FGT sum rule, as in \cite{AFSS-EPL}) shows qualitatively 
the opposite, conventional behavior in the case of the overdoped sample \cite{AFSS-SNS}.

\section{Experimental results}
Reflectivity spectra, recorded in the range 30-25000~cm$^{-1}$ for 15 temperatures between 
300~K and 10~K, are reported elsewhere \cite{AFSS-PRL}. An example is recalled in Fig.1,  
for selected temperatures, up to 3000~cm$^{-1}$, for the 
underdoped sample ($T_c$=70~K). Most of the change with temperature occurs 
below 1000~cm$^{-1}$. However, the difference between the spectra at 10~K and 100~K extends up 
to 5000~cm$^{-1}$, displaying a reflectivity increase in the superconducting state of 0.3\%, in this range, 
to be compared to the 4\% increase below 1000~cm$^{-1}$ (see inset of fig.1).

\begin{figure}[ht]
\epsfig{figure=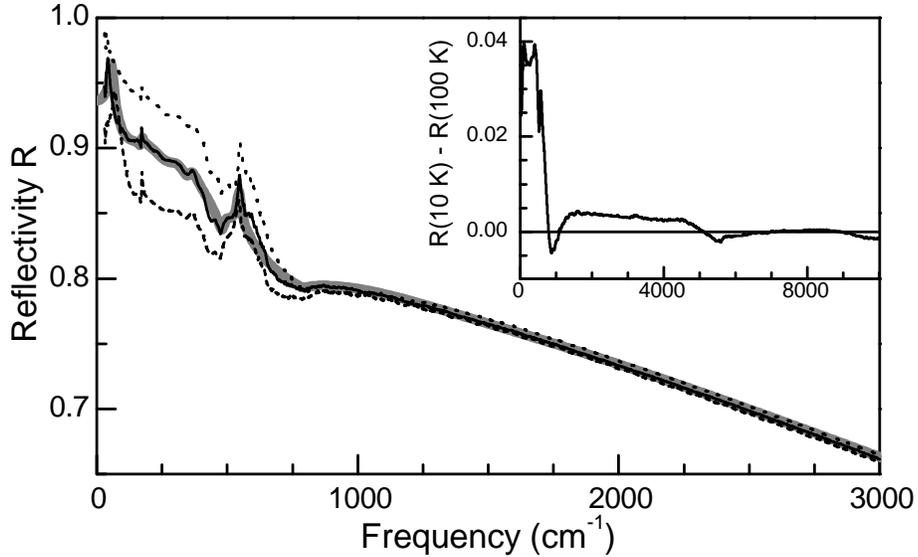, width=\linewidth}
\caption{Reflectivity spectra  in the 0,3000~cm$^{-1}$ range at three temperatures: 
200~K: dashed, 100~K: solid, 10~K: dotted line. The thick gray line is the fitted 
spectrum at 100~K. Inset: difference R(10~K)- R(100~K) between the reflectivity at 
10~K and 100~K.}
\label{reffig1}
\end{figure}

A standard thin film fitting procedure 
was applied in order to derive the optical conductivity \cite{AFSS-PRL,Santander-thèse}. 
The  fit is accurate enough so as to reproduce within less than 0.1~\%, the small reflectivity 
difference between 100 and 10~K. This illustrates the accuracy of our data {\it and} analysis.
A Kramers-Kronig based argument shows that such a small difference in reflectivity over a given frequency range $\Delta \omega$ 
results into changes in conductivity extending over 
a frequency range ~2-3~$\Delta \omega$ \cite{Santander-thèse}. The changes of spectral weights 
over an unusually large frequency 
range (possibly up to 16000~cm$^{-1}$ as discussed further), arises from this general argument.
 Without the  ability of resolving 0.1~\% in the reflectivity, the results to be described 
 further could not possibly be  firmly established.
The optical conductivity is shown in Fig.2, in the same frequency range and the same 
temperatures as in Fig.1.

\begin{figure}[ht]
\epsfig{figure=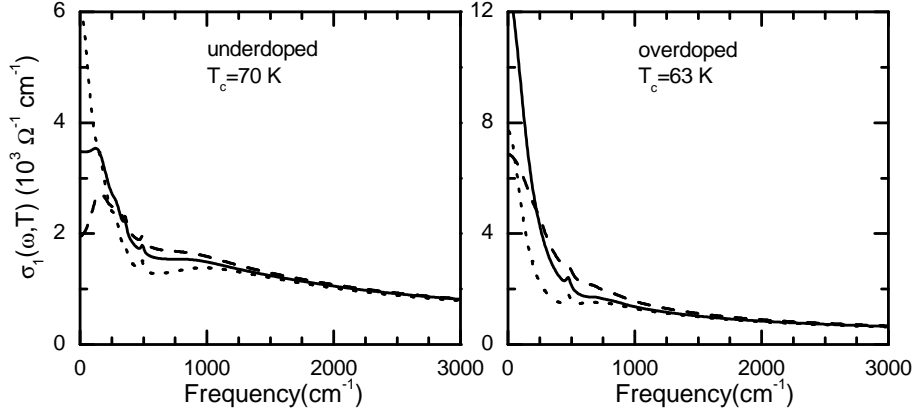, width=\linewidth}
\caption{Real part $\sigma_{1}(\omega, T)$ of the optical conductivity, for the 
underdoped and overdoped (for comparison) samples. Temperatures are the same as in fig.1. 
Note that, in the 
former, a very large Drude-like contribution persists in the superconducting state.}
\label{reffig2}
\end{figure}

\section{Kinetic energy} 
From the optical conductivity, we compute the spectral weight or partial sum rule defined in Eq.1. 
We show in Fig3-a and -b the temperature variation of the spectral 
weight for selected  cut-off frequencies $\Omega$ ranging from 500 to 8000~cm$^{-1}$, 
 both for the underdoped sample (500, 1000, 8000~cm$^{-1}$), 
and for the overdoped sample 
(8000~cm$^{-1}$).
For sake of clarity, we have normalized each spectral weight to its value at 300K.
The integration starting from 
$0^+$ , in the superconducting state, does not include the 
spectral weight $W_s$ of the $\delta$ function. We have added below $T_c$ the weight 
of the superfluid, 
using the input parameter determined by the best fit \cite{superfluid}. 
We thus obtain, below $T_c$, the data represented by the 
open symbols in Fig.3, hereafter referred to as the {\it total} spectral weight.  
We have carefully worked out the accuracy of the data by using various 
sets of fitting parameters (mainly the superfluid  weight and
the Drude contribution) and by estimating the uncertainty on each experimental point accordingly. 
The size of symbols in Fig.3-b has been adjusted so as to represent the uncertainty.

\begin{figure}[ht]
\epsfig{figure=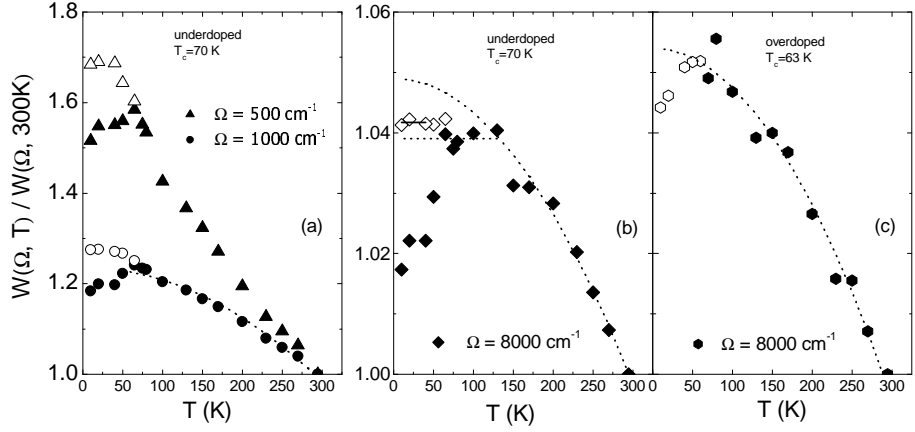, width=\linewidth, clip=}
\caption{Partial sum rule shown for underdoped (a) and (b), and overdoped (c) samples, for selected 
cutoff frequencies. Full symbols represent the spectral weight, integrated from $0^+$, hence 
without the superfluid contribution. Open symbols include (below $T_c$) the superfluid weight. 
Fig3-b and -c represent the intraband spectral weight, hence $-E_k$, as a 
function of temperature.
The dotted lines are $T^2$ best fits to the normal state data.}
\label{reffig3}
\end{figure}

We see in Fig.3-a  that, when integrating up to 
1000~cm$^{-1}$, the normal state spectral weight above $T_c$ 
follows a $T^2$ increase, as indicated by a  dotted line. The  total spectral weight 
exceeds significantly this normal state spectral weight. 
Considering 1000~cm$^{-1}$ as a conventional energy scale as discussed in the introduction, we 
are led to conclude that the energy scale over which one must extend the integration, in order 
to retrieve the spectral weight of the $\delta$ 
function, exceeds significantly any conventional scale. 
The excess in total spectral weight in the superconducting state, with respect to the normal state,
 is present throughout the conduction band, i.e. up to 
typically 1~eV=8000~cm$^{-1}$ (fig.3-b). In this range, the spectral weight follows a $T^2$ behavior down to 130~K, 
 then levels off.  Between 
the normal (horizontal dotted line) and superconducting state (horizontal solid line), we observe 
 a small but definite increase. 
  Equation 2 yields then the change of kinetic energy 
 associated with this change of spectral weight.
We find $\Delta E_{k} = - (0.5 \pm 0.3$)~meV per copper site, thus consistent with our early data 
\cite{AFSS-EPL} and confirming that the onset of superconductivity in the underdoped Bi-2212 
sample is associated with a decrease of kinetic energy.
In contrast, the overdoped sample displays a very clear decrease of spectral weight when performing the 
integration up to 8000~cm$^{-1}$ (fig.3-c), i.e. an {\it increase} of kinetic energy. This is 
the conventional BCS behavior \cite{VDMEk}. 
In order to make the connection with our earlier work, we now translate briefly these results 
 in terms of the FGT sum rule. The FGT sum rule is fulfilled for the overdoped 
sample over a conventional range  (1000~cm$^{-1}$) \cite{AFSS-EPL}. Conversely, it is {\it not satisfied}
 up to 8000~cm$^{-1}$ in the underdoped sample. 
 With our refined analysis, we can set a lower limit of 
~10000~cm$^{-1}$ and an upper limit of ~16000~cm$^{-1}$ for the energy scale required to satisfy the FGT 
sum rule. Recent work also shows a much larger energy scale for the exhaustion of 
the FGT sum rule in underdoped YBCO, as compared to the optimal doping \cite{CCHomes}.
 Such large energy scale is difficult to reconcile with a simple phonon mechanism, and seems 
 therefore to call for a different excitation spectrum or more elaborate mechanisms.
 
   Condensation by  kinetic energy saving \cite{NormanPepin2,Eksav} has been considered by several groups, 
   sometimes with reasonable quantitative agreement  \cite{NormanPepin2}.
    Other models have been discussed in order to  account for these data 
    \cite{Hirschetal}.

\section{Summary}
 In conclusion, we have shown that according to the doping level in Bi-2212, the onset of 
  superconductivity is associated with a decrease of kinetic energy in the underdoped material, whereas 
  the
  overdoped material exhibits an increase of kinetic energy. Such a drastic difference  confirms that 
  the superconducting state in the underdoped regime of   cuprates is  unconventional.  

\begin{acknowledgments}
The authors wish to thank M. Norman for numerous discussions and illuminating remarks, 
and Z. Konstantinovic, Z.Z. Li and H. Raffy for kindly providing 
high quality films.
\end{acknowledgments}


\begin{chapthebibliography}{1}
\bibitem{superfluid} When a large Drude-like contribution is present, the standard derivation of 
the superfluid weight from either $\epsilon_1 (\omega)$ versus $\omega^{-2}$ as in ref. \cite{AFSS-EPL} or 
$\omega \sigma_2 (\omega)$ versus $\omega$ is biased. The best estimate comes out to be the 
input value of our fit \cite{MN}.
\end{chapthebibliography}
\end{document}